\documentclass[runningheads]{llncs}
\usepackage[T1]{fontenc}
\usepackage{amssymb} 
\usepackage{amsmath}
\usepackage{mathrsfs}
\usepackage{hyperref}

\usepackage{tikz}
\usepackage{float}
\usepackage{adjustbox}
\usepackage{multirow}
\usepackage{makecell}
\usetikzlibrary{arrows.meta}
\usepackage{graphicx}
\graphicspath{graphics/}
\begin{document}
\title{STV Audit Graphs: \\ A Visual Tool to Measure Election Stability}
\titlerunning{STV Audit Graphs}
%
%
\author{Edouard Heitzmann\inst{1}\orcidID{0009-0004-6410-7832}}
\authorrunning{E. Heitzmann}
%
\institute{Department of Mathematics, University of Colorado in Boulder \email{edouard.heitzmann@colorado.edu}}
\maketitle              
\begin{abstract}
The Single Transferable Vote (STV) is an algorithmic election rule. Round by round, a profile of ranked-choice ballots is reinterpreted to determine which decision to make next, and candidates are seated or eliminated until a full winner set emerges. This algorithmic nature makes STV theoretically more brittle than other election rules: uncertainty about an early round of the election might percolate irreversibly into the rest. For this reason, a generalized non-trivial Risk-Limiting Audit (RLA) framework has remained elusive for STV. Such a framework must concoct a set of null hypotheses, or assertions, whose rejection would bound the probability that the outcome of the election was incorrectly reported. We present audit graphs as a solution to design the assertions needed for RLAs of arbitrary STV elections, as well as quantitatively describe the uncertainty (or lack thereof) of their outcomes. These audit graphs explore election paths that are ``close'' to the recorded one by considering the alternative decisions the STV algorithm might have made if a small number of ballots were perturbed.

\end{abstract}

\section{Introduction.}
The Single Transferable Vote (STV) is an algorithmic election rule. 
Round by round, a profile of ranked-choice ballots is re-interpreted to determine which decision to make next, and candidates are seated or eliminated until a full winner set arises.

This algorithmic nature makes STV theoretically more brittle than other election rules. 
If an early decision in the algorithm is contingent on a sharp margin, a small perturbation of the ballot profile might have a ``butterfly effect'' on the election outcome. 
For example, running the STV ruleset on an arbitrarily large partial sample of the profile might systematically produce the wrong winner(s) \cite{icelandSamplingWinnersRanked2024}. 
It is further possible for a voter to disadvantage their preferred candidate by casting a ballot supporting them, creating a so-called voting ``paradox'' \cite{mccuneMonotonicityAnomaliesScottish2024}.

The severity of these theoretical pathologies is perhaps tempered by their empirical rarity in real-world data.
The same references point out that running the STV ruleset on a partial sample of the profile usually gives a good approximation of the actual outcome in real data, and the frequency of any voting paradox in recorded profiles is below 6\% \cite{icelandSamplingWinnersRanked2024,mccuneMonotonicityAnomaliesScottish2024}.

We introduce audit graphs as a tool to quantitatively resolve the tension between theoretical brittleness and empirical stability of the STV rule for a given profile.
By directly considering alternative election sequences that would result from the reversal of contingent sharp margins, they are able to detect or rule out the possibility of one of these pathological instabilities.
They further offer a path to compute desirable lower bounds for the margin of victory of an STV contest, as well as an intuitive and visual way to understand the macroscopic support patterns of a ranked-choice profile.

Eponymously, audit graphs also provide a natural and general framework to design assertion-based ballot-comparison Risk-Limiting Audits (RLAs) for STV elections. 
RLAs are statistical tests to verify the reported outcome of an election by sampling the paper ballots while bounding the risk of a type I error (miscertification).
There is already a generalized framework to create RLAs for single-winner STV elections \cite{blomRAIRERiskLimitingAudits2019b}, and there exist partial frameworks to audit multi-winner STV elections that fall within a given set of patterns \cite{blom3+SeatRiskLimiting2026}. 
Audit graphs unify and generalize these frameworks: a sufficient test is to reject the hypothesis that the true election path can escape a preconvened audit graph.

\renewcommand{\subset}{\subseteq}

\newcommand{\PPPP}{\mathscr{P}}

\newcommand{\CCC}{\mathcal{C}}
\newcommand{\EEE}{\mathcal{E}}
\newcommand{\HHH}{\mathcal{H}}
\newcommand{\WWW}{\mathcal{W}}
\newcommand{\VS}{\mathcal{VS}}
\newcommand{\SSS}{\mathcal{S}}
\newcommand{\LLL}{\mathcal{L}}
\newcommand{\OOO}{\mathcal{O}}

\newcommand{\MPT}{\mathsf{MPT}}

\newcommand{\ZZ}{\mathbb{Z}}
\newcommand{\RR}{\mathbb{R}}
\newcommand{\NN}{\mathbb{N}}

\newcommand{\tv}{\tilde{v}}
\newcommand{\tu}{\tilde{u}}
\newcommand{\ve}{\vec{e}}
\newcommand{\ov}{\overline{v}}
\newcommand{\ou}{\overline{u}}
\newcommand{\oO}{\overline{\Omega}}
\let\vec\relax\newcommand{\vec}[1]{\mathaccent"017E\relax{#1}}

\section{Construction.}
Intuitively, the vertices of an audit graph represent possible rounds the STV algorithm might tabulate, and the edges encode possible decisions it could make in each round.

We fix a \emph{universal set of candidates} $\CCC = \{0,1,\ldots, C-1\}$ for some $C\in \ZZ^+$ as well as a \emph{number of seats} $m\in \ZZ^+$. 
These two quantities are sufficient to define the \emph{universal audit graph} $\Omega = \Omega(C,m)$, which will induce the plausible graphs we use to quantify election stability.
\subsection{The Universal Audit Graph.}
A \emph{pre-vertex} $\tv$ is an equivalence class of vertices of the audit graph consisting of two pieces of information:
\begin{enumerate}
	\item a set of \emph{hopeful candidates} $\HHH = \HHH(\tv)\subset \CCC$.
	\item a list of \emph{seated winners} $\WWW = \WWW(\tv)\subset\CCC$ satisfying $\WWW\cap\HHH = \varnothing$, $|\WWW\cup\HHH|\geq m$, and $|\WWW|\leq m$. We deliberately define $\WWW$ as an (ordered) list so that vertices may remember the order their winners were seated in.
\end{enumerate}
We supply the set of pre-vertices with a weak partial ordering: $\tu\leq\tv$ when $\HHH(\tu)\subset\HHH(\tv)$ and $\WWW(\tv)$ is a prefix of the list $\WWW(\tu)$ -- i.e. it is a contiguous initial portion of the list. 
We read $\tu\leq \tv$ by saying $\tu$ is \emph{descended from} $\tv$.

To turn a pre-vertex $\tv$ into a vertex $v$ of the audit graph $\Omega$, we equip it with a third piece of data:
\begin{enumerate}
	\item[3.] a \emph{seating chart} for the pre-vertex $\tv$ is a map $\lambda = \lambda_v \colon \WWW(\tv)\to\Omega$ mapping each winner $w$ to her \emph{seating vertex} $\lambda(w)\in\Omega$. This map must satisfy:
		\begin{enumerate}
		    \item $w$ was hopeful when she was seated: $w\in \HHH(\lambda(w))$.
		    \item the seating chart of $v$ only points to vertices it is descended from: $v\leq \lambda_v(w)$ for all $w\in \WWW(v)$.
		    \item the seating chart respects the chronology of $\WWW$: $\lambda(\WWW[i]) \leq \lambda(\WWW[i-1])$ for all $1\leq i < |\WWW(v)|$.
		    \item seating charts are shared with ancestors: for all $w \in \WWW(v)$, for all $w'\in \WWW(\lambda_v(w))\cap \WWW(v)$, $\lambda_v(w')=\lambda_{\lambda_v(w)}(w')$.
		\end{enumerate}
\end{enumerate}
The purpose of keeping track of where each winner was seated is to allow the vertex to re-compute its tallies given a perturbation of the ballots.
Informally, STV computes these tallies by assigning a ``transfer value'' $\tau(w)$ to each winner $w\in\WWW(v)$, which indicates what fraction of their weight each ballot is allowed to keep if it counted towards $w$'s quota.
This transfer value depends on $w$'s tally at the time of her seating, which in turn depends on which other candidates were still standing at that point. 
The seating chart $\lambda$ tracks all this information in a deliberately rich way so that $v$'s tallies can be recomputed using a purely local parameterization of the ballots.

Note that each vertex is descended from itself, so consecutive seated winners might share the same seating vertex (indicating that they were seated in the same round of the election).
There exists at least one seating chart for each pre-vertex $\tv$. The vertex set of $\Omega$ consists of all possible pairs $v = (\tv, \lambda_v)$ of pre-vertices equipped with a seating chart.
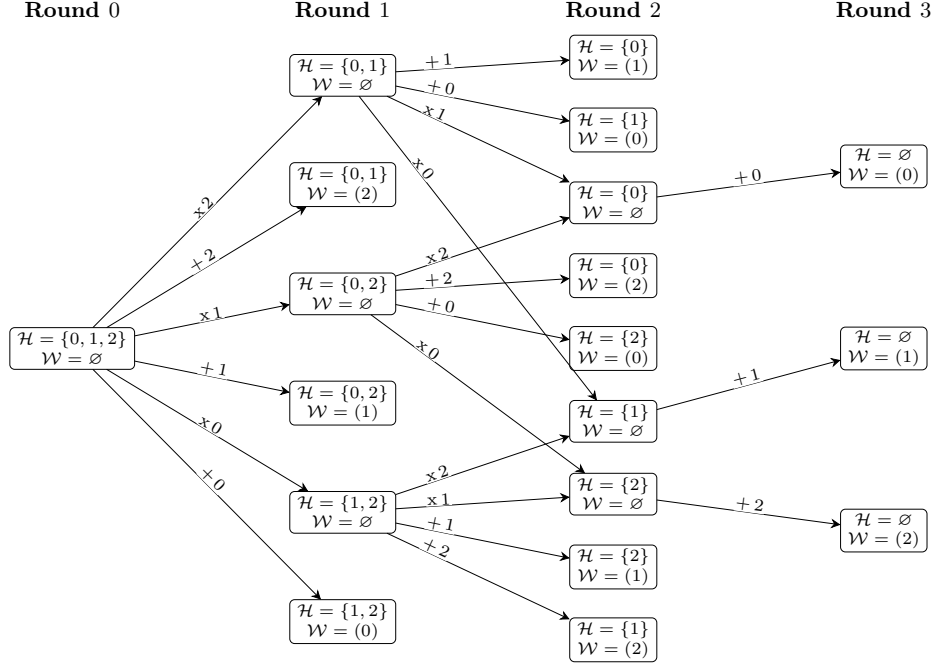
\begin{figure}[h]
\centering

\setlength{\abovecaptionskip}{2pt}
\setlength{\belowcaptionskip}{0pt}

\begin{adjustbox}{max width=\linewidth,center}
\begin{tikzpicture}[
    x=3.9cm,
    y=1.05cm,
    vertex/.style={
        draw,
        rounded corners=2pt,
        fill=white,
        align=center,
        inner xsep=3pt,
        inner ysep=1.5pt,
        outer sep=0pt,
        font=\scriptsize
    },
    edge/.style={
        -{Stealth[length=4pt,width=4pt]},
        thin
    },
    edge label/.style={
        fill=white,
        inner sep=0.5pt,
        outer sep=0pt,
        font=\scriptsize
    },
    layer label/.style={
    font=\small\bfseries,
    anchor=south,
    inner sep=0pt,
    outer sep=0pt,
    yshift=6pt
}
]

\node[layer label] at (0,4.35) {Round $0$};
\node[layer label] at (1,4.35) {Round $1$};
\node[layer label] at (2,4.35) {Round $2$};
\node[layer label] at (3,4.35) {Round $3$};

\node[vertex] (root) at (0,0)
    {$\mathcal H=\{0,1,2\}$\\$\mathcal W=\varnothing$};

\node[vertex] (n01) at (1,3.75)
    {$\mathcal H=\{0,1\}$\\$\mathcal W=\varnothing$};

\node[vertex] (s2) at (1,2.25)
    {$\mathcal H=\{0,1\}$\\$\mathcal W=(2)$};

\node[vertex] (n02) at (1,0.75)
    {$\mathcal H=\{0,2\}$\\$\mathcal W=\varnothing$};

\node[vertex] (s1) at (1,-0.75)
    {$\mathcal H=\{0,2\}$\\$\mathcal W=(1)$};

\node[vertex] (n12) at (1,-2.25)
    {$\mathcal H=\{1,2\}$\\$\mathcal W=\varnothing$};

\node[vertex] (s0) at (1,-3.75)
    {$\mathcal H=\{1,2\}$\\$\mathcal W=(0)$};

\node[vertex] (h0w1) at (2,4)
    {$\mathcal H=\{0\}$\\$\mathcal W=(1)$};

\node[vertex] (h1w0) at (2,3)
    {$\mathcal H=\{1\}$\\$\mathcal W=(0)$};

\node[vertex] (h0n) at (2,2)
    {$\mathcal H=\{0\}$\\$\mathcal W=\varnothing$};

\node[vertex] (h0w2) at (2,1)
    {$\mathcal H=\{0\}$\\$\mathcal W=(2)$};

\node[vertex] (h2w0) at (2,0)
    {$\mathcal H=\{2\}$\\$\mathcal W=(0)$};

\node[vertex] (h1n) at (2,-1)
    {$\mathcal H=\{1\}$\\$\mathcal W=\varnothing$};

\node[vertex] (h2n) at (2,-2)
    {$\mathcal H=\{2\}$\\$\mathcal W=\varnothing$};

\node[vertex] (h2w1) at (2,-3)
    {$\mathcal H=\{2\}$\\$\mathcal W=(1)$};

\node[vertex] (h1w2) at (2,-4)
    {$\mathcal H=\{1\}$\\$\mathcal W=(2)$};

\node[vertex] (z0) at (3,2.5)
    {$\mathcal H=\varnothing$\\$\mathcal W=(0)$};

\node[vertex] (z1) at (3,0)
    {$\mathcal H=\varnothing$\\$\mathcal W=(1)$};

\node[vertex] (z2) at (3,-2.5)
    {$\mathcal H=\varnothing$\\$\mathcal W=(2)$};

\draw[edge]
    (root)
    --
    node[edge label,sloped,above] {$\mathrm{x}\,2$}
    (n01);

\draw[edge]
    (root)
    --
    node[edge label,sloped,above] {$+\,2$}
    (s2);

\draw[edge]
    (root)
    --
    node[edge label,sloped,above] {$\mathrm{x}\,1$}
    (n02);

\draw[edge]
    (root)
    --
    node[edge label,sloped,above] {$+\,1$}
    (s1);

\draw[edge]
    (root)
    --
    node[edge label,sloped,above] {$\mathrm{x}\,0$}
    (n12);

\draw[edge]
    (root)
    --
    node[edge label,sloped,above] {$+\,0$}
    (s0);

\draw[edge]
    (n01)
    --
    node[edge label,sloped,above,pos=0.25] {$+\,1$}
    (h0w1);

\draw[edge]
    (n01)
    --
    node[edge label,sloped,above,pos=0.25] {$+\,0$}
    (h1w0);

\draw[edge]
    (n01)
    --
    node[edge label,sloped,above,pos=0.25] {$\mathrm{x}\,1$}
    (h0n);

\draw[edge]
    (n01)
    --
    node[edge label,sloped,above,pos=0.25] {$\mathrm{x}\,0$}
    (h1n);

\draw[edge]
    (n02)
    --
    node[edge label,sloped,above,pos=0.25] {$+\,2$}
    (h0w2);

\draw[edge]
    (n02)
    --
    node[edge label,sloped,above,pos=0.25] {$+\,0$}
    (h2w0);

\draw[edge]
    (n02)
    --
    node[edge label,sloped,above,pos=0.25] {$\mathrm{x}\,2$}
    (h0n);

\draw[edge]
    (n02)
    --
    node[edge label,sloped,above,pos=0.25] {$\mathrm{x}\,0$}
    (h2n);

\draw[edge]
    (n12)
    --
    node[edge label,sloped,above,pos=0.25] {$+\,1$}
    (h2w1);

\draw[edge]
    (n12)
    --
    node[edge label,sloped,above,pos=0.25] {$+\,2$}
    (h1w2);

\draw[edge]
    (n12)
    --
    node[edge label,sloped,above,pos=0.25] {$\mathrm{x}\,1$}
    (h2n);

\draw[edge]
    (n12)
    --
    node[edge label,sloped,above,pos=0.25] {$\mathrm{x}\,2$}
    (h1n);

\draw[edge]
    (h0n)
    --
    node[edge label,sloped,above] {$+\,0$}
    (z0);

\draw[edge]
    (h1n)
    --
    node[edge label,sloped,above] {$+\,1$}
    (z1);

\draw[edge]
    (h2n)
    --
    node[edge label,sloped,above] {$+\,2$}
    (z2);

\end{tikzpicture}
\end{adjustbox}

\caption{The universal audit graph $\Omega(3,1)$.}
\end{figure}

We draw an oriented edge $\ve = vu$ from vertex $v$ to vertex $u$ in $\Omega$ when $\tu\leq\tv$ and either one of two things happen:
\begin{enumerate}
    \item $|\WWW(u)|= |\WWW(v)|\neq m$ and $|\HHH(v)-\HHH(u)|=1$. 
	    In this case, we call the edge $\ve$ an \emph{elimination edge}, and we identify the singleton element $l\in\HHH(v)-\HHH(u)$ as the \emph{loser} corresponding to this edge.\medskip
    \item $\HHH(v)-\HHH(u) = \WWW(u)-\WWW(v)\neq\varnothing$ and $\lambda_u(w)=v$ for all $w\in \WWW(u)-\WWW(v)$. 
	    In this case, we call the edge $\ve$ a \emph{seating edge}, and we call the candidates $w_1, \ldots, w_k \in \WWW(u)-\WWW(v)$ the \emph{winners} corresponding to this edge. When $k>1$, we say these winners are \emph{simultaneous}.
\end{enumerate}

It will further be useful to define the \emph{degree} of a vertex $v$ as the size of its winner set, i.e. $d(v)= |\WWW(v)|$, and the \emph{Round} of a vertex $v$ as $\text{Round}(v) = C-|\HHH(v)|$.

With the edge set defined as above, the graph $\Omega$ is layered with respect to Round number, since edges $\ve = u v$ must satisfy $\text{Round}(u)<\text{Round}(v)$. 
The graph $\Omega$ is also rooted: the root is the vertex $(\HHH,\WWW)=(\CCC,\varnothing)$ equipped with the trivial seating chart.

\subsection{Plausible Graphs.}\label{sec:plausible}

To carve out a subgraph from $\Omega$ which is actually descriptive of a given election, our approach will be to consider paths which might have been followed by the STV algorithm but for a small perturbation of the ballots.
We keep a lot of the formal details of this process for Appendix \ref{sec:appendix}.

Informally, a \emph{ranking} is a repetition-free list of elements from $\CCC$, and a \emph{profile} is a list of rankings. 
We use the letter $N$ to indicate the number of rankings in the profile, and we denote the set of all length-$N$ profiles on $C$ candidates as $\PPPP = \PPPP(C, N)$.

Then, a \emph{tally function} is a map $T\colon \PPPP\times V(\Omega)\to \RR^C$ that interprets a given profile in the context of a vertex of the audit graph to assign tallies to each of the $C$ candidates\footnote{\label{note1}The tally function and election rule also have implicit access to the number $m$ of seats, which is fixed throughout.
}. 

Informally, this tally function uses the seating chart of a vertex to compute transfer values for each winner using their seating vertex tallies. 
This means that winners might be seated early or late, and their transfer values might be deflated or inflated as a result. The details are in Equation (\ref{eq:transfer}) of Appendix \ref{sec:appendix}.

When the tally function and underlying profile $P\in\PPPP$ are fixed, the \emph{tallied universal graph} is the graph $\overline{\Omega}=\{\big(v, T(P,v)\big)\colon v\in\Omega)\}$ consisting of all vertices in $\Omega$ labelled with their numerical tallies.
The edge set of $\overline{\Omega}$ is identical to the edge set of $\Omega$.

Finally, an \emph{election rule} is a map $\EEE\colon \overline{\Omega}\to \overline{\Omega}$ which takes each tallied vertex to a vertex in its oriented closed neighborhood\footnotemark[1] -- i.e. the map must satisfy $\overline{v}\EEE(\overline{v})\in E(\overline{\Omega})$ or $\EEE(\overline{v}) = \overline{v}$ for all $\overline{v}\in\overline{\Omega}$.

There are many variants of the STV ruleset, but a generic version, called the Weighted Inclusive Gregory Rule, or WIGM, would fix a quantity called quota at $q = \lfloor N/(m+1)\rfloor +1$, and then use the following election rule:
\begin{itemize}
	\item If $\overline{v}$ is a leaf, map it to itself.
	\item Otherwise, if a hopeful candidate has a tally greater than or equal to quota, follow the seating edge where the winners are all candidates $w\in \HHH(\overline{v})$ with $T(w)\geq q$. If there are multiple winners, order them from highest to lowest tally.
	\item Otherwise, if $|\HHH(\ov)|=m-d(\ov)$, follow the seating edge where the winners are $\HHH(\ov)$.
	\item Otherwise, follow the elimination edge where the loser is the candidate $l\in \HHH(\ov)$ with the lowest tally, breaking ties in a well-defined way.
\end{itemize}

For each non-leaf $\ov\in \overline{\Omega}$, we call the unique edge $\ov\EEE(\ov)$ that the election path would follow the \emph{natural edge} of $\ov$.
There is a unique \emph{natural path} connecting the root of $\oO$ to a distinguished \emph{natural leaf} of $\oO$ consisting only of natural edges: this is the path corresponding to the reported election outcome.

In the context where we are uncertain about the accuracy of the rankings in the profile $P$, we might also wish to consider paths that are ``close'' to the natural path in some sense.
One way to do this is to fix a \emph{Margin of Insecurity} $M\in \ZZ^+$, and to call an edge $\ov\,\ou$ \emph{plausible} when it could be natural if the tallies of $\ov$ were shifted by an $L^1$ distance of at most $M$.

It will further be helpful vocabulary to define the \emph{plausibility threshold} of an edge $\ve\in E(\oO)$ leaving from $\ov$ as the smallest $L^1$ distance from the tallies of $\ov$ to a list of tallies that would make $\ve$ natural.
So an edge is plausible exactly when its plausibility threshold is no more than $M$.

Then, we construct the \emph{plausible graph} $G=G(\oO, M)$ corresponding to this margin by including any vertex that could be reached from the root of $\oO$ via a sequence of plausible edges, and taking the induced subgraph.

It is helpful to think of $G$ as the set of election paths that might result from giving an adversary a budget of $M/2$ ballots to modify in each round of the election.
Whatever alternative election path they might achieve will be contained within the plausible graph.

This construction is deliberately generous to the adversary, as we reset their budget each round: they might perturb one set of $M/2$ ballots in the first round of the election, at which point we reset the profile to its starting point, and the adversary may choose a (possibly different) set of $M/2$ ballots to perturb in the next round.
We choose to be generous in this way to make plausibility a purely local property. This makes the graphs easier to construct, and auditable via a set of simple, vertex-local null hypotheses.

Later, our strategy to turn a plausible graph $G$ into an assertion-based RLA will be to verify a set of assertions which amount to ``the natural path is contained within $G$.''
Of course, this is only helpful if all election outcomes contained in $G$ are consistent with the reported outcome. 
So we say $G$ is \emph{coherent} if for all leaves $\ov\in G$, the winner list $\WWW(\ov)$ is a permutation of the winner list of the natural leaf of $G$.
To keep the audit sample size small, it will be desirable to make $M$ as large as possible while keeping $G$ coherent.

\begin{figure}
    \centering
    \includegraphics[width=\textwidth]{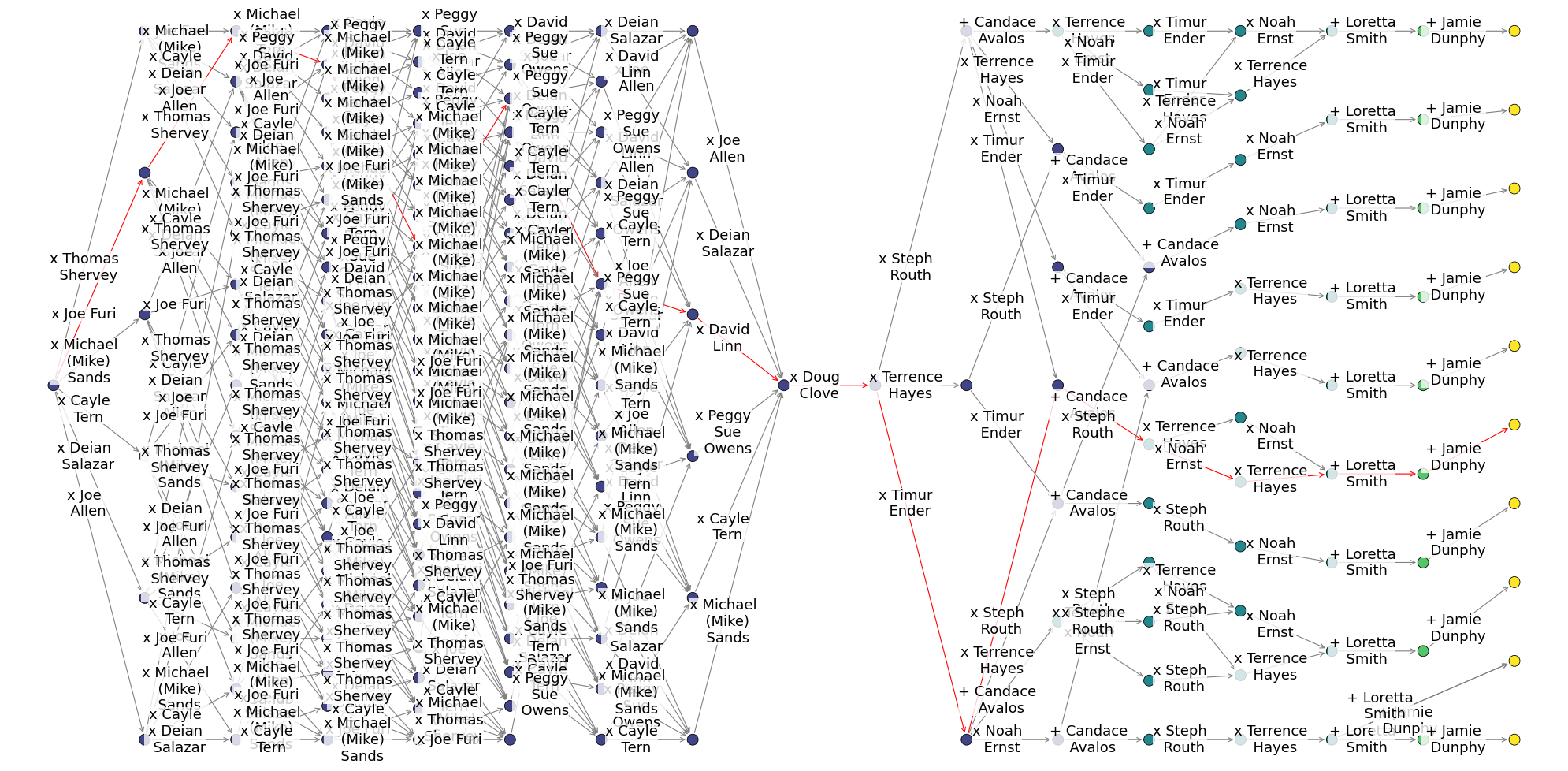}
    \caption{A coherent and plausible graph for a 2024 3-seat STV election in Portland, OR. This election had $C=16$ candidates, $m=3$ seats, $N=42,678$ ballots, and the graph was constructed with a margin of $M=671$. The vertices are color-coded by degree, and the natural path is highlighted in red.}
    \label{fig:portland_d1}
\end{figure}

\subsection{Shortcuts for Larger Elections.}\label{sec:shortcut}

With larger candidate counts, the computation of a full plausible graph becomes computationally intractable.
The early layers of the audit graphs are especially prone to combinatorial explosion, as there tends to be a large set of ``weak candidates'' whose elimination is plausible at any stage of the election. 

This pattern already emerges in Figure \ref{fig:portland_d1}, where layers 0 through 7 of the graph contain $111/175$ of the vertices of the graph, and it is further accentuated for larger elections, where the first half of the graph's layers routinely contain more than $90\%$ of the vertices.

We can leverage some systematic patterns from real-world data to mitigate this combinatorial explosion.
For example, strong candidates tend to start strong and remain so throughout in real world elections, unlike the pathological examples which imagine ``dark horse'' winners always on the brink of elimination.
In this setting, it is possible to abridge the early layers of a plausible graph by using a ``batch elimination'' which rigorously justifies the simultaneous elimination of all candidates in a designated weak set. 

To justify a batch elimination, we fix a Margin of Insecurity $M$, and we distinguish two disjoint subsets $\VS, \SSS\subset \CCC$ of \emph{Very Strong} and \emph{Strong} candidates.

The (possibly empty) set $\VS$ consists of those candidates who make quota by more than $M$ votes on initial preferences, or upon receiving transfers from other very strong candidates. 
In other words, these are the candidates who will be seated before anything else happens in the plausible graph: any path from the root to a leaf of the plausible graph must initially follow a sequence of seating edges corresponding to very strong candidates.
We call the possibly multiple vertices where all very strong candidates are seated and no other candidates were eliminated or seated the \emph{base vertices} of the batch eliminations.

The set $\SSS$ of strong candidates can be chosen arbitrarily, but it induces a \emph{Weak Set} $\LLL\subseteq \CCC$ on each base vertex, and in practice it will be desirable to pick $\SSS$ so as to make $\LLL$ as large as possible.
We say a candidate $l\in \CCC$ is \emph{weak} with respect to $\SSS$ in the base vertex $\ov$ if her \emph{Maximum Possible Tally} ($\MPT$) is lower than the smallest tally of a strong candidate according to $\ov$.
Informally, we count any ranking which mentions $l$ before any strong candidate in $\SSS$ as giving its full weight in $\ov$ to $\MPT(l)$.

After computing the weak set $\LLL$, we are left with a set $\OOO = \CCC - (\VS\cup\SSS\cup\LLL)$ of \emph{Other} candidates which fell into none of the previous categories.

If we verify that none of the strong candidates can possibly make quota before all of the weak candidates are eliminated, the above definitions guarantee that any plausible path must at some point visit a vertex descended from a base vertex whose hopeful set contains all of $\SSS$ and some subset $\OOO'\subseteq \OOO$.
So we may abridge the process of plausible graph construction by seeding it with the set of all such vertices, rather than the root vertex.

\begin{figure}
    \centering
    \includegraphics[width=\textwidth]{"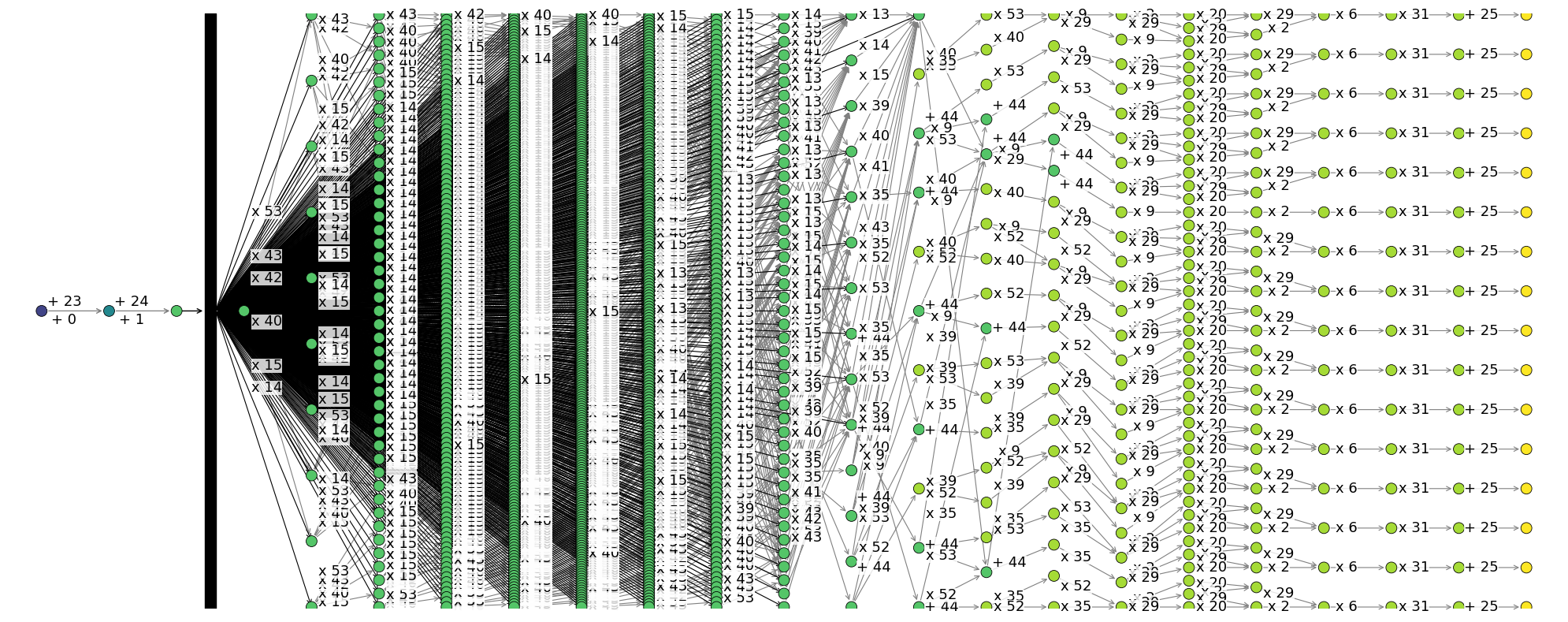"}
\caption{Plausible and coherent graph for a 6-seat STV election for Federal Senate in Victoria, Australia. For this graph, $N=4,101,762$, $M=23,000$, $C = 65$, $|\VS|=4$, $|\SSS|=9$, $|\LLL|=42$, and $|\OOO|=10$, meaning there were $2^{10}=1024$ seeded vertices after the batch elimination, which is indicated in the graph by a vertical black bar.}
    \label{fig:vic}
\end{figure}

To encode the requirement that no strong candidate makes quota until all weak candidates are eliminated at the stage of graph construction, it is sufficient to verify that the tallies of all strong candidates are at least $M$ below quota in the vertex $\ou$ descended from the base vertex $\ov$ with $\HHH(\ou) = \SSS$ and $\lambda_{\ou} = \lambda_{\ov}$.
This requirement, like the weakness of all $l\in\LLL$, is contingent on a critical margin which is purely vertex-local to either $\ov$ or $\ou$, meaning they can be formulated into assertions that are auditable by the methods in \S\ref{sec:mismatch} or \S\ref{sec:delta}.

\newcommand{\pG}{\partial G}

\section{Applications.}
\subsection{Margin of Victory Bounds.}\label{sec:mov}
The \emph{Margin of Victory} (MoV) of an election is the smallest number of ballots that must be modified in order to change the outcome of the election. 
Computing the MoV is useful for election reporting -- voters are bound to be curious how close the election was -- and also for election auditing (cf. \S\ref{sec:mismatch}).

STV is among the most difficult election rules to find a margin for. 
Even approximating the MoV is known to be NP-hard 
for STV; indeed, just deciding whether MoV$=1$ is NP-hard \cite{xiaComputingMarginVictory2012}.

Computing a lower bound on the MoV is more productive. 
There is already substantial work on this problem using a branch-and-bound Mixed-Integer Non-Linear Program (MINLP) \cite{blomAdvancesSTVMargin2026}, and such a bound may still be used to formulate an RLA for the election \cite{ekDoingMoreLess2026}.
Audit graphs offer an alternative jumping-off point to the computation of these bounds, which might be used in tandem with the MINLP approach to tighten them.

The approach follows the intuition of \S\ref{sec:plausible}: a coherent $M$-plausible audit graph $G$ is a witness that manipulating fewer than $M/2$ ballots can not change the outcome of the election, \emph{provided that} such a manipulation can not shift the tallies in any vertex of $G$ by an $L^1$ distance of more than $M$.
This provision holds as long as the tally function satisfies some notion of ``conservation of ballot mass:'' the sum of a single ballot's contribution to all tallies is at most $1$.

This is a principle that the STV tabulation algorithm was specifically encoded to preserve.
Ballots start the election with a weight of $1$, and transfer values are ordinarily computed to live in $[0,1]$.
The only difference in the context of audit graph construction is the risk of negative transfer values: when a winner's tally is contained in $[q-M, q)$, the edge corresponding to her seating will be considered plausible, even though she did not make quota.
We call such a winner \emph{insecure}.

When a vertex contains two insecure winners, the possibility of ballots transferring through two negative transfer values imperils the conservation of ballot mass.
In such a situation the positive ballot mass outgoing from the second insecure winner's seating, which corresponds to ballots which transferred through the negative transfer values of both insecure winners, has the ability to freely displace ballot mass from one side to the other of any downstream critical margin.

To account for this, we define the \emph{double-negative mass} $D_w$ outgoing from a winner $w \in \WWW(v)$ as the amount of positive ballot mass transferring out of her seating, and we say an $M$-plausible audit graph $\overline{G}$ is $\emph{weakly secure}$ if the plausibility threshold of each edge $e= \ov\,\ou$ pointing from a vertex $\ov\in G$ to a vertex $\ou\notin G$ is greater than $M + 2\sum_{w\in \WWW(\ov)} D_w$.
\begin{theorem}\label{thm:reasonable}
As long as $M\leq q$, a coherent and weakly secure $M$-plausible graph $G$ witnesses that the MoV of an election is greater than $M/2$.
\end{theorem}

It bears emphasis that the construction of plausible graphs is not optimized for MoV computation. 
In particular, the generosity described in \S\ref{sec:plausible} means that a plausible path contained in a graph $G$ might not be realizable by a perturbation of $\leq M/2$ ballots.
The path might need to change one set of $M/2$ ballots in one round of the election, and another disjoint set in the next.

A shrewder MoV-bounding approach would be to use an audit graph to detect the most plausible election path leading to an alternative winner set, then run the MINLP algorithm of \cite{blomAdvancesSTVMargin2026} on the prefix of that path leading to its edge with the largest plausibility threshold. 
Then, replace said threshold with the program's output, and repeat the process with the next-most plausible alternative path.

Imperfect as they are, plausible graphs can still be close to sharp in some cases.
For example, the plausible graph in Figure \ref{fig:vic} witnesses a lower bound $\text{MoV}\geq 11,500$ for its election, which closes the gap with a known upper bound of $\text{MoV}\leq 16,021$ computed using the ConcreteSTV algorithm of \cite{blomAdvancesSTVMargin2026}.


\subsection{Edge-Local Mismatch Audits.}\label{sec:mismatch}
The (global) mismatch-based approach generates an RLA by verifying that fewer than a MoV worth of votes were reported incorrectly \cite{ekDoingMoreLess2026}. 
This is a so-called ``ballot-comparison'' audit: we have access to a digital \emph{Cast Vote Record} (CVR) indicating how each vote was interpreted by the tabulation, as well as a physical \emph{Manual Vote Record} (MVR) representing a paper trail which can be compared to the record by an Auditor.
The data input into the RLA test process is a sequence of CVR-MVR pairs which are compared for mismatches.

Audit graphs allow for a new subtlety in mismatch-audit design, by giving the auditing process a way of filtering out ``irrelevant'' mismatches. 
For example, in a single-seat contest with three candidates $\CCC = \{a, b, c\}$, if there is a separate test process already verifying that $c$ can not win the election, then a mismatch between a ballot reported as $(a,b)$ in the CVR but sampled as $(a, c)$ in the MVR can safely be ignored; either way, the ballot will count towards candidate $a$'s tally when it matters.

To formalize this, given a fixed plausible and weakly secure graph $G\subset \oO$, we call an edge $\ve = \ou\,\ov\in E(\oO)$ an \emph{escape} edge if $\ou\in G$ but $\ov\notin G$. 
We use the notation $\partial G$ to denote the set of all escape edges of $G$.

We define our \emph{global null hypothesis} as $H_{\star}=$ ``the natural election path under the MVR escapes the graph $G$'', and for each $\ve\in\pG$ we define a \emph{local null hypothesis} $H_{\ve}=$ ``the edge $\ve$ is natural.''
Under the SHANGRLA framework, a valid test of $H_{\star}$ at level $\alpha$ is to reject each of the $H_{\ve}$ at level $\alpha$, since $H_{\star}\subset \bigcup_{\ve\in\pG} H_{\ve}$. In particular, there is no need to adjust for multiplicity -- cf. \S4.1 of \cite{starkSetsHalfaverageNulls2020}.

To reject a local null $H_{\ve}$ where $\ve = \ou\, \ov$, it is always sufficient to verify that a given affine function of the tallies in $\ou$ which was recorded as positive in the CVR stays positive in the MVR. We call this affine function the \emph{critical margin} controlling the plausibility threshold of $\ve$ -- in practice, it can take one of three forms, depending on $\ve$:
\begin{itemize}
	\item The critical margin can be a \emph{candidate-to-candidate} margin, of the form $M_{cl} = T(c) - T(l)$, where $c, l \in \HHH(\ou)$ are two distinguished candidates.
	\item Alternatively, the critical margin can be a \emph{candidate-below-quota} or \emph{candidate-above-quota} margin, $M_{qc} = q-T(c)$ or $M_{cq} = T(c) -q$, where $c\in\HHH(\ou)$ is a distinguished candidate.
\end{itemize}
For example, if $\ve$ is a seating edge corresponding to winners $S\subset \HHH(\ou)$, a valid critical margin is the candidate-below-quota margin $M_{qc}$ corresponding to the candidate $c\in S$ with the lowest tally according to $\ou$. 
If $\ve$ is an elimination edge corresponding to $c\in \HHH(\ou)$, a valid critical margin might be the candidate-to-candidate margin $M_{cl}$ where $l\in \HHH(\ou)$ is the hopeful candidate with the lowest recorded tally in $\ou$, or alternatively the candidate-above-quota margin $M_{wq}$ where $w\in \HHH(\ou)$ is the candidate with the highest tally in excess of quota.

Depending on which of these margins we are working with, we can drastically reduce the dimension of the parameter space established in Equation (\ref{eq:parameters}) of Appendix \ref{sec:appendix} from $(|\HHH(\ou)|+1)2^{d(\ou)}$ to either $3\cdot 2^{d(\ou)}$ or $2^{d(\ou)+1}$.
For example, in a candidate-to-candidate margin $M_{cl}$, any parameter $t_{S;o}$ where $o\notin \{c,l\}$ has an identical impact, as long as $S$ is the same -- so they may all be treated as a single ``reduced parameter.''

Then at the level of a local null $H_{\ve}$, a CVR-MVR pair should only be considered a mismatch if their interpretation in this reduced parameter space differs.
To reject $H_{\ve}$, it is sufficient to check that there are not enough of these mismatches to reverse the critical margin computed by the CVR.
This in turn can be done via any of the standard test supermartingales integrated in the SHANGRLA framework, such as ALPHA or COBRA \cite{starkALPHAAuditThat2023c,spertusCOBRAComparisonOptimalBetting2023}.
We use the latter for the results in \S\ref{sec:results}.

The advantage of considering mismatches at the local rather than global level is a dilution of the overall noise level between the CVR and MVR.
Assuming the noise is not adversarial, most CVR-MVR discrepancies will only affect a fraction of the escape edges in $\pG$, leading to a more efficient test process.

When assigning the plausibility threshold for an escape edge downstream from at least two insecure seatings, we should keep in mind, in view of the proof of Theorem \ref{thm:reasonable} in Appendix \ref{sec:proof}, that we should take away twice the double-negative mass of all insecure winners from the tallied vertex's recorded critical margin. 
This is generally a negligible cost for real-world profiles; cf. Section \ref{sec:results}.
\subsection{Delta-Method Audits.}\label{sec:delta}

We might also consider alternative test processes to reject the local null hypotheses $H_{\ve}$ of the previous section.
In particular, a test process which successfully uses the information contained in each mismatch -- which candidate the mismatch helps or hurts -- is bound to be more efficient in the non-adversarial setting.

However, the non-linearity of the tally function in Equation (\ref{eq:tally}) of Appendix \ref{sec:appendix} means that the standard test supermartingales can't be applied off the shelf.
This is because the direct impact of a sampled mismatch on the critical margin depends not only on the mismatch itself, but also the sampling history of other mismatches.
For example, sampling the same mismatch repeatedly can have a self-reinforcing impact on the margin.
We must adapt our test processes to these non-linear margins.

One such test uses the delta method to construct confidence intervals for each critical margin, and rejects the local null if this interval is positive. 
This involves bounding the variances and covariances of the sample means of each of the reduced parameters from \S\ref{sec:mismatch} and multiplying by an appropriate gradient matrix to obtain a variance for the critical margin, from which we build a confidence interval -- cf. \cite{heitzmannGraphBasedAuditsMeek2026a} for details.

This test process uses a standard Wald Confidence Interval, and as such it requires a fixed sample size -- it does not share the arbitrary stopping times of test supermartingales.
On the other hand, it is more resilient to noise than tests like mismatch-based audits, because in the regime of low noise it replaces degenerate sample variances with worst-case stochastic bounds which are more sensitive to the sample size than to the (low) number of sampled mismatches.
But this also means that the Delta method performs worse for edges leaving higher-degree vertices, since these variance bounds are scaled by the number of parameters in Equation (\ref{eq:parameters}) to give the variance of the critical margin, and that number of parameters grows exponentially in the degree of the base vertex.

The Delta method also need not worry about weak security. Nowhere does it rely on the availability of bounds on the MoV of the election (not even local bounds), and the possibility of a double-negative ballot transfer is accounted for by the gradient vector which multiplies the covariance matrix that is central to the method.

In summary, the delta method has some advantages and drawbacks compared to mismatch-based audits. 
It should not be taken as a final ``best practice'' test process to audit the assertions that audit graphs give rise to, but rather as a first pass attempt at using the information contained within a CVR-MVR comparison to build a more efficient test process, which will be improved (indeed, already has been improved) in due time.
We show some results comparing the performance of the mismatch and delta methods in the next section. 
\subsection{Results.}\label{sec:results}
We constructed audit graphs and generated RLAs for a representative cross-section of STV elections from across the world.
For Scottish City Council and New South Wales (NSW) Local Government elections, we randomly selected one profile to audit for each of $m=2$, $3$, $4$, and $5$-seat elections. 
Elsewhere, we aimed to represent multiple scales of STV elections, and we prioritized recent elections in municipalities that still use STV today.

We recorded, for each of those elections, the maximal Margin of Insecurity (MoI) $M$ for which a coherent and plausible graph could be constructed, and the Average Sampling Number (ASN) required for a successful audit.
We artificially noised these election CVRs with $2\%$ mismatches, following a standard noising procedure established in the Appendix of \cite{blomRAIRERiskLimitingAudits2019b}. We used a risk level of $\alpha = 5\%$ for each of these audits.

For the Mismatch method, which has arbitrary stopping times, we computed the ASN as the average sample size of 10 audits of the election, assuming all 10 certified with a sample size of less than $N/2$. 
For the Delta method, we defined the ASN as the smallest sample size which would lead to a successful audit in at least 9 out of 10 randomly seeded trials.
\begin{table}[h]
\centering
\caption{Margins and Sample Sizes for Graph-Based Audits of STV Elections.}
\label{tab:results}

\resizebox{\textwidth}{!}{%
\begin{tabular}{@{}clrrrrrrr@{}}
\hline
Category & Election & Year & $C$ & $m$ & $N$
& $M$
& Mismatch ASN
& Delta ASN \\
\hline
\hline

\multirow[c]{4}{*}{\makecell[c]{Scottish\\Local\\Elections}}
& Sgire Nan Loch
& 2022 & 4 & 2 & 851
& 141 (17\%)
& 66 (7.8\%)
& 56 (6.6\%) \\
\cline{2-9}

& Bearsden North
& 2017 & 6 & 3 & 7,035
& 87 (1.2\%)
& X
& 2,345 (33.3\%) \\
\cline{2-9}

& Ardrossan and Arran
& 2012 & 10 & 4 & 5,590
& 75 (1.3\%)
& X
& 1,118 (20\%) \\
\cline{2-9}

& Saltcoats and Stevenston
& 2022 & 8 & 5 & 6,506
& 148 (2.3\%)
& X
& 650 (10\%) \\
\hline\hline

\multirow[c]{4}{*}{\makecell[c]{NSW\\Local\\Elections}}
& Shellharbour Ward D
& 2024 & 6 & 2 & 11,600
& 1,933 (17\%)
& 51.7 (0.45\%)
& 46 (0.4\%) \\
\cline{2-9}

& Cumberland Greystanes
& 2017 & 13 & 3 & 19,422
& 21 (0.11\%)
& X
& X \\
\cline{2-9}

& Wollongong Ward 3
& 2021 & 16 & 4 & 40,706
& 2,200\textsuperscript{2} (5\%)
& X
& 1,017 (2.5\%) \\
\cline{2-9}

& Penrith South
& 2024 & 15 & 5 & 37,778
& 600 (1.6\%)
& X
& 3,777 (10\%) \\
\hline\hline

\multirow[c]{2}{*}{Albany, CA}
& City Council
& 2022 & 5 & 3 & 7,159
& 129 (2\%)
& X
& 715 (10\%) \\
\cline{2-9}

& City Council
& 2024 & 5 & 3 & 8,007
& 192 (2.4\%)
& X
& 533 (6.7\%) \\
\hline\hline

\multirow[c]{4}{*}{\makecell[c]{Portland, OR\\City Council}}
& District 1
& 2024 & 16 & 3 & 42,678
& 671 (1.6\%)
& X
& 711 (1.7\%) \\
\cline{2-9}

& District 2
& 2024 & 22 & 3 & 77,024
& 2,739\textsuperscript{1} (3.6\%)
& 575.5 (0.75\%)
& 308 (0.4\%) \\
\cline{2-9}

& District 3
& 2024 & 30 & 3 & 84,356
& 300 (0.36\%)
& X
& 4,217 (5\%) \\
\cline{2-9}

& District 4
& 2024 & 30 & 3 & 76,566
& 200 (0.26\%)
& X
& 7,656 (10\%) \\
\hline\hline

\multirow[c]{4}{*}{Minneapolis, MN}
& Parks Board
& 2025 & 8 & 3 & 113,348
& 3,209 (2.8\%)
& X
& 453 (0.4\%) \\
\cline{2-9}

& Parks Board
& 2021 & 7 & 3 & 106,650
& 1,111 (1\%)
& X
& 1,777 (1.7\%) \\
\cline{2-9}

& Board of Taxation
& 2025 & 3 & 2 & 106,682
& 8,467 (7.9\%)
& 425.3 (0.4\%)
& 133 (0.12\%) \\
\cline{2-9}

& Board of Taxation
& 2021 & 4 & 2 & 95,625
& 5,406 (5.7\%)
& 1,528.4 (1.6\%)
& 159 (0.17\%) \\
\hline\hline

\multirow[c]{8}{*}{\makecell[c]{Australian\\Senate}}
& ACT
& 2025 & 14 & 2 & 293,474
& 15,000 (5\%)
& 301.1 (0.1\%)
& 293 (0.1\%) \\
\cline{2-9}

& Northern Territory
& 2025 & 17 & 2 & 106,807
& 1,500 (1.4\%)
& 2,371.9 (2.2\%)
& 1,068 (1\%) \\
\cline{2-9}

& Victoria
& 2025 & 65 & 6 & 4,101,762
& 20,000\textsuperscript{1} (0.5\%)
& X
& 20,508 (0.5\%) \\
\cline{2-9}

& New South Wales
& 2025 & 56 & 6 & 4,986,832 & 12,000\textsuperscript{1} (0.2\%)
& X & 49,868 (1\%) \\
\cline{2-9}

& Tasmania
& 2025 & 33 & 6 & 371,790
& 5,000\textsuperscript{1} (1\%)
& 3,253.9 (1.3\%)
& 1,487 (0.4\%) \\
\cline{2-9}

& South Australia
& 2025 & 40 & 6 & 1,164,072 &
7,000\textsuperscript{1} (0.6\%) & X & 11,640 (1\%) \\
\cline{2-9}

& Queensland
& 2025 & 56 & 6 & 3,224,436 & 30,000\textsuperscript{1} (1\%)
& X & 8,061 (0.25\%) \\
\cline{2-9}

& Western Australia
& 2025 & 49 & 6 & 1,622,016 & 2,850\textsuperscript{1} (0.2\%)
& X & 108,134 (7\%) \\
\hline

\end{tabular}%
}

\vspace{2pt}

\begin{minipage}{\textwidth}
\footnotesize
\textsuperscript{1} The audit graph was constructed with a batch elimination.\quad
\textsuperscript{2} For Wollongong Ward 3, \(M=500\) was used for audits.
\end{minipage}
\vspace{-12pt}
\end{table}

There are two potential failure points for graph-based STV audits -- namely graph construction and assertion testing -- and the data provides insight into both.

When graph construction fails, it is usually because of combinatorial explosion of the early layers of the graph for elections with $C\geq 20$ candidates. 
The batch elimination shortcut of \S\ref{sec:shortcut} works to circumvent this problem in some cases, but it does not apply in situations where a strong candidate is close to making a quota on initial preferences, as in Wollongong or Portland District 3.
The latter is factually one of the most stable in Portland -- the three winners are close to making a solid coalition -- but because batch elimination does not apply, the only remedy is to construct a full plausible graph with a much smaller margin $M$.

It is generally possible to construct a plausible graph by pushing $M$ low enough, at the cost of a higher ASN for the audit it generates.
This tradeoff is best illustrated by the Cumberland Greystanes election, which was simply a close race: the last round was decided by a margin of 21 out of the nearly 20,000 votes cast.
This closeness was responsible for the only outright failure of the Delta method.

The Local Mismatch Audit ASN was only comparable to the Delta-method Audit in profiles where the diluted margin $M/N$ was large, as is predicted by the theory of the COBRA test process in \S 3 of \cite{spertusCOBRAComparisonOptimalBetting2023}.
In fairness to the Mismatch Audit, its success rates are much higher when noise levels are closer to the real-world measured rates (typically $<0.2\%$).
We chose a deliberately higher noise level so that our results might demonstrate reliability as well as effectiveness.

Of all the audit graphs we constructed, only three contained vertices with at least two insecure winners -- namely Ardrossan and Arran, Portland District 2, and Minneapolis Board of Taxation 2025.
Of these three, only one graph had to pay a cost for weak security: the maximal MoI that led to a coherent and weakly secure graph for Ardrossan and Arran had a MoI of $73$, paying the weight of $2$ votes for weak security.
Since the Local Mismatch method was unsuccessful for that profile in either case, and the Delta Method does not require a weakly secure graph, this cost had no bearing on our results.

These results indicate how STV RLAs might further be improved -- e.g. devising more shortcuts to avoid full-graph computation -- but they are also cause for optimism. In particular, this is the first published RLA of an Australian 6-seat Senate election, the largest use of STV in the world.

Replication code for Table \ref{tab:results} is available in the \href{https://github.com/EdouardHeitzmann/audit_graphs.git}{\underline{companion repository}} for this paper.

\section{Further Work.}

The most immediate improvement of the graph-based RLA framework would be to adapt a test supermartingale to the non-linear margins of STV. 
This would allow for more efficient and flexible audits with arbitrary stopping times.

The shortcut of \S\ref{sec:shortcut} might also be extended to deal with cases where a strong candidate is seated in the middle of a batch elimination.

Another line to look into is the usage of audit graphs for partial audits of STV elections. 
Conspicuously absent from \S\ref{sec:results} are the 9-seat elections of Cambridge, MA: this is because their higher seat counts make the margin for the last seat systematically razor sharp.
In such a case audit graphs can still formulate an audit for the first 8 out of 9 seats.
The construction of a ``partially coherent'' graph might also allow for the study of per-candidate Margins of Victory.

More broadly, there is a vast canyon separating our understanding of how chaotic STV elections can be in the worst case and how well-behaved they tend to be in reality.
This paper puts a rope bridge over this canyon, but there is still a lot of work to be done in resolving the tension between theoretical brittleness and empirical stability of the Single Transferable Vote.

\begin{credits}
\subsubsection{\ackname} I am deeply grateful to Damjan Vukcevic, Michelle Blom, Peter Stuckey, and Vanessa Teague for their hospitality and generosity in inviting me to visit Monash University for 6 weeks in June of 2026. I further thank Moon Duchin and the Data and Democracy Lab for funding my research for a formative semester in the Fall of 2025, and I thank Andrew Conway for publishing Australian CVR data as part of the ConcreteSTV project.
\end{credits}

\appendix

\newcommand{\fpv}{\mathsf{fpv}}
\newcommand{\pre}{\mathsf{pre}}

\newcommand{\PPP}{\mathcal{P}}

\section{The STV Tally Function.}\label{sec:appendix}
It will be useful in time to have a formal vertex-local parameterization for the STV tally function. We still work with a fixed candidate set $\CCC = [C]$, a seat number $m\in\ZZ^+$, and a number of ballots $N\in\ZZ^+$.
As before, we define the \emph{quota} as $q = \lfloor N/(m+1)\rfloor +1$. 

Formally, a \emph{ranking} on candidates $\HHH\subset \CCC$ is a monomorphism $r\colon [k]\hookrightarrow \HHH$ for some $k\in \NN$. We allow $k=0$, in which case $r=\varnothing$ is the empty ranking. 
We denote the set of all rankings on a subset of candidates $\HHH\subset \CCC$ as $R_{\HHH}$. Then a \emph{profile} is formally a map $P\colon [N]\to R_{\CCC}$.

Note that when $\HHH'\subseteq \HHH$, we have $R_{\HHH'}\subseteq R_{\HHH}$. To map a ranking in the reverse direction, we define the projection map $\pi\colon R_{\HHH}\to R_{\HHH'}$ as $\pi(r)= r\circ \iota$, where $\iota\colon [k']\hookrightarrow [k]$ is the unique increasing monomorphism satisfying $r\circ\iota([k'])\subset \HHH'$ while maximizing $k'\in\NN$. This $k'$ might be $0$, in which case the projection $\pi(r)$ is empty, i.e. the ranking $r$ is ``exhausted'' in $\HHH'$.

Given a vertex $v = (\tv, \lambda)\in\Omega(C, m)$, the \emph{projection} $\pi_v$ \emph{onto the vertex} $v$ is just the projection map from $R_{\CCC}$ to $R_{\HHH(v)}$. In that setting we also define the \emph{first-place vote map} $\fpv_v \colon R_{\CCC}\to \CCC\cup\{-1\}$ as follows:
\[
	\fpv_v(r)=\begin{cases}
		-1 & \text{if } \pi_v(r)=\varnothing;\\
		\pi_v(r)(0) & \text{otherwise.}
	\end{cases}
\]
This in turn allows us to define the \emph{transfer prefix map} $\pre_v\colon R_{\CCC}\to \PPP(\WWW(v))$ as the map which takes each ranking to the set of winners that ranking would have transferred through according to $v$'s seating chart:
\[
	\pre_v(r) = \{w\in\WWW(v)\colon \fpv_{\lambda_v(w)}(r)=w\}.
\]
We're now equipped to discuss the set of parameters we use to encode the tally function at the level of a vertex $v$. These are the \emph{profile totals transferring to} $c\in\CCC$ \emph{through} $S\subset \WWW(v)$ \emph{according to} $v$:
\begin{equation}
	t_{S;c}(P, v) = |\{j\in [N]\colon \pre_v\big(P(j)\big)=S \text{ and } \fpv_v\big(P(j)\big)=c\}|.
	\label{eq:parameters}
\end{equation}
We allow $S=\varnothing$ (votes that transfer through no winners before getting to their current first preference in $v$) as well as $c=-1$ (votes that indicate no preference in the current hopeful set).

Finally we can recursively define the \emph{tallies} $T_c$ of candidates $c\in \CCC$ and the \emph{transfer values} of the winners $w\in \WWW(v)$:
\begin{align}
	T_c(P,v) &= \sum_{S\subset \WWW(v)}\left(\prod_{w\in S} \tau_w (P,v)\right) t_{S;c}(P,v) 
	\label{eq:tally} \\
	\tau_w(P,v) &= \Big(T_w\big(P, \lambda_v(w)\big)-q\Big)\Big/ T_w\big(P, \lambda_v(w)\big).
	\label{eq:transfer}
\end{align}
This is a well-defined recursion, because each winner tally $T_w(P, \lambda_v(w))$ only depends on the transfer values of previously seated winners.
In particular, when $w_1, w_2$ are simultaneous winners, any total transferring through both winners must be zero, because the $\fpv$ map is well-defined -- so we do not need to know the transfer value of $w_2$ to compute the tally of $w_1$.

However, it is possible for the denominator in Equation (\ref{eq:transfer}) to be zero.
In such a case we call the vertex $v$ \emph{degenerate}, and the tally function can be defined arbitrarily for $v$.
This won't matter, since the plausible graph construction of \S\ref{sec:plausible} will never need to visit a degenerate vertex so long as $M$ is given a reasonable value -- certainly $M<q$ suffices, in view of Theorem \ref{thm:reasonable}.

Even though the transfer value equations in (\ref{eq:transfer}) nominally need access to the profile totals $t_{S;w}(P,\lambda_v(w))$ of a previous vertex, these can be recuperated from the vertex-local parameters in (\ref{eq:parameters}) as follows:
\[
	t_{S;w}\big(P,\lambda_v(w)\big) = \sum_{c\in\HHH(v) \cup\{-1\}}\sum_{\substack{
			S'\subset \WWW(v) \\
			S'\cap \WWW(\lambda_v(w)) = S \\
			w\in S'
	}} t_{S';c}(P,v).
\]
So the tally function can be considered to be a function exclusively of the parameters in Equation (\ref{eq:parameters}), of which there are $D=(|\HHH(v)|+1)2^{|\WWW(v)|}$. In particular, every ballot in the profile can faithfully be represented as a unit vector in the hypercube $\{0,1\}^D$.

\newcommand{\zr}{\big|}
\newcommand{\vx}{\vec{x}}
\newcommand{\vy}{\vec{y}}

In order to formally parameterize the double-negative mass $D_w$ outgoing from the seating of a winner $w$, it will help to establish another piece of vocabulary.
Given a profile $P\in \PPPP(C,N)$ and a vertex $v\in \Omega(C,m)$, we define the \emph{ranking mass map} $\mu(P, v)\colon R_{\CCC} \to\RR$ as follows:
\[
	\mu\Big(P, v\Big)(r) = |P^{-1}[\{r\}]|\cdot \prod \left\{ \tau_{w}(P, v)\colon \fpv_{\lambda_{v} (w)}(r) =w \right\}.
\]
Plainly, $\mu(P, v)(r)$ measures the total mass of all rankings of type $r$ in the profile $P$ at the stage of vertex $v$: it multiplies the quantity of such rankings in $P$ by their weight after transferring through all the winners indicated by the seating chart of $v$.

With this new definition, given a winner $w\in\WWW(v)$, we use the notation $A_w = \fpv^{-1}_{\lambda_v(w)}[\{w\}]\subseteq R_\CCC$ to indicate the subset of rankings that would count for $w$ in her seating vertex, and we formally define the quantity $D_w(P,v)$ as follows:
\[
	D_w(P,v) = \min\big(0,\tau_w (P,v)\big) \sum_{r\in A_w}\min\left( 0, \mu\Big(P, \lambda_v(w)\Big)(r) \right)
\]

\section{Proof of Theorem 1.}\label{sec:proof}
We fix a vertex $v \in \Omega(C,m)$, as well as two profiles $P,P' \in \PPPP(C,N)$ that differ in fewer than $M/2<q/2$ ballots, in the sense that
\[
	d(P,P') := |\{j\in [N]\colon P(j)\neq P'(j)\}|\leq \frac{M}{2}.
\]
We further fix an edge $\ve = v u$ outgoing from $v$, which we assume is plausible according to $P$ and lies directly on the natural path of the election according to $P'$.
We make the following claim:
\begin{equation}
	||\mu(P, v) - \mu(P', v)||_1 \leq M +2\sum_{w\in \WWW(v)}D_w(P,v).\label{eq:claim}
\end{equation}

This also acts as a bound on the $L^1$ distance between the tallies of $v$ according to $P$ versus $P'$, because said tallies are linear combinations of the components of the ranking mass maps according to their respective profiles. 
This in turn gives us a bound on the plausibility threshold of $\ve$ according to $P$: since $\ve$ is natural according to $P'$, its plausibility threshold is $0$ according to that profile, and replacing $P'$ with $P$ can increase this threshold by no more than the $L^1$ distance between the tallies of its base vertex according to both profiles.

The theorem will follow from Claim (\ref{eq:claim}), because it implies that the natural path of $P'$ will be contained in any $M$-plausible and weakly secure graph constructed according to $P$.
If $P'$ represents the ``ground truth'' profile MVR, this guarantees that the winner set $\WWW'$ of its natural leaf will be the same as the winner set $\WWW$ of the natural leaf of the CVR profile $P$.
This is because the natural leaf of $P$ is contained in any $M$-plausible weakly secure graph constructed according to $P$, which by the Claim in (\ref{eq:claim}) must also contain the entire natural path of $P'$; since the Theorem assumes there exists such a graph that is coherent, both must have the same winner set.

To prove the Claim (\ref{eq:claim}), we go by strong induction on the length $L$ of the segment of the natural path of $P'$ leading to $\ve$. 

For the base case, when $L=0$ and $v$ is the root of $\Omega(C,m)$, we note that when $P$ and $P'$ only differ by a single ballot, i.e. $d(P,P')=1$, then the $L^1$ distance in the left-hand side of (\ref{eq:claim}) is at most $2$, since that ballot can only change any given entry of the ranking mass map $\mu(P,v)$ by $1$, and it can only change at most two entries.
Hence by the triangle inequality, we have
\[
||\mu(P,v) - \mu(P', v)||_1 \leq 2 d(P, P') \leq M.
\]
For the inductive step, we assume the claim is true as long as the natural path of $P'$ leading to $v$ is of length at most $L-1$, and we assume $\ve$ is the $L^{th}$ edge on that path.
In that case we replace $v$ with $u$ in the left-hand side of Claim (\ref{eq:claim}): we must show
\begin{equation}
	||\mu(P, u) - \mu(P', u)||_1 \leq M +2\sum_{w\in \WWW(u)}D_w(P,u).\label{eq:is}
\end{equation}
We further break this up into cases according to the type of edge that $\ve = v u$ encodes. If it is an elimination edge, the inductive step is immediate, because such an edge does not change any of the entries of the ranking mass map: plainly,
\[
	\mu(P,v)=\mu(P,u) \; \text{ and } \; \mu(P',v)=\mu(P',u).
\]
On the other hand, when $\ve$ is a seating edge, let $S(e)\subseteq \HHH(v)$ be the set of winners who are simultaneously seated in this edge.
For each $w\in S(\ve)$, we re-use the notation $A_w = \fpv^{-1}_v[\{w\}]=\{r\in R_\CCC \colon \fpv_v(r)=w\}\subseteq R_\CCC$ to indicate all rankings that sat with $w$ when she was seated by $\ve$.
This induces a partition of $R_\CCC$:
\[
	R_\CCC = O \cup \bigcup_{w\in S(e)}A_w, \; \text{ where } \; O = R_\CCC - \bigcup_{w\in S(e)} A_w.
\]
Because these sets are disjoint, we can break up the $L^1$ distance in the left-hand side of Claim (\ref{eq:is}) into fragments covering the restriction of the ranking mass maps to each element of the partition:
\begin{multline}
	||\mu(P,u) - \mu(P',u)||_1 = \\ ||\mu(P,u)\big|_O - \mu(P',u)\big|_O||_1+\sum_{w\in S(\ve)} ||\mu(P,u)\big|_{A_w} - \mu(P',u)\big|_{A_w}||_1\label{eq:partition}
\end{multline}
Clearly, the restriction of the ranking mass map to $O$ is the same applied to $u$ as it is to $v$, because rankings that did not sit with a winner in $v$ do not get reweighed in $u$. 
We will show, for every $w \in S(e)$, that 
\begin{equation}
	||\mu(P,u)\big|_{A_w}-\mu(P',u)\big|_{A_w}||_1\leq ||\mu(P,v)\big|_{A_w}-\mu(P',v)\big|_{A_w}||_1 + 2D_w(P,u).\label{eq:critical_step}
\end{equation}
To show how the inductive step follows from (\ref{eq:critical_step}), it will be useful to use the shorthand $\mu'(u)=\mu(P',u)$, $\mu'(v)=\mu(P', v)$, $\mu(u)=\mu(P,u)$, and $\mu(v)=\mu(P,v)$.
Then we see:
\begin{align*}
	||\mu(u)&-\mu'(u)||_1 = ||\mu(u)\zr_O-\mu'(u)\zr_O||_1 + \sum_{w\in S(\ve)}||\mu(u)\zr_{A_w}- \mu'(u)\zr_{A_w}||_1 \tag{by (\ref{eq:partition})}\\
			     &= ||\mu(v)\zr_O-\mu'(v)\zr_O||_1 + \sum_{w\in S(\ve)}||\mu(u)\zr_{A_w}- \mu'(u)\zr_{A_w}||_1 \tag{$\mu(u)\zr_O=\mu(v)\zr_O$}\\
			     &\leq ||\mu(v)\zr_O-\mu'(v)\zr_O||_1 + \sum_{w\in S(\ve)}\left(||\mu(v)\zr_{A_w}- \mu'(v)\zr_{A_w}||_1 + 2D_w(P,u)\right) \tag{by (\ref{eq:critical_step})}\\
			     & = ||\mu(v)-\mu'(v)||_1 + 2\sum_{w\in S(\ve)} D_w(P,u) \tag{same as (\ref{eq:partition})}\\
			     & \leq \left( M + 2\sum_{w\in \WWW(v)} D_w(P, v)\right) + 2\sum_{w\in S(\ve)} D_w(P,u) \tag{inductive hypothesis}\\
			     & = M + 2\sum_{w\in \WWW(u)}D_w(P, u) \tag{$\WWW(u)=\WWW(v)\cup S(\ve)$}
\end{align*}
This would prove the inductive step, assuming we justify the critical step in (\ref{eq:critical_step}). To do this, it will be useful to set some further notation, given a fixed winner $w\in S(\ve)$:
\begin{itemize}
	\item $\vx=\mu(P,v)\zr_{A_w}$ is the restriction of the ranking mass map to $A_w$ for the profile $P$.
	\item $\vy=\mu(P',v)\zr_{A_w}$ is the same for $P'$.
	\item $X = \sum_{r\in A_w} \vx(r) = T_w(P,v)$ is the tally of $w$ at the time of her seating according to $P$.
	\item $Y = \sum_{r\in A_w} \vy(r) = T_w(P',v)$ is the same thing for $P'$. 
		Because the seating of $w$ in $v$ was natural according to $P'$, we can assume (without loss of generality) that $Y\geq q$; the only alternative is that $u$ is a leaf, in which case the transfer value of $w$ is inconsequential, because there will be no more escape edges after $\ve$.
		This naturality also implies all entries of $\vy$ are non-negative, hence $Y=||\vy||_1$.
	\item $\alpha = \tau_w(P, u) = 1 - q/X$ is the transfer value of $w$ according to $P$.
	\item $\beta = \tau_w(P',u)=1-q/Y$ is the same for $P'$.
	\item $D_w = D_w(P,u)$ is shorthand because the profile and vertex are clear (in particular, note $D_w(P', u)=0$ because the seating of $w$ is natural in $u$).
\end{itemize}
With this fixed notation, the inequality we must show for (\ref{eq:critical_step}) looks as follows:
\[
||\alpha \vx - \beta \vy||_1 \leq ||\vx -\vy||_1 +2D_w.
\]
We split the proof of this inequality into cases according to whether or not $w$ is secure according to $P$, i.e. whether $X<q$ or $X\geq q$.\medskip

\noindent \underline{Case 1:} $X\geq q$.\medskip

In this case we note that $\alpha\in[0,1]$, hence $q/X=(1-\alpha)\in[0,1]$, and we see
\begin{align*}
	||\alpha \vx - \beta \vy ||_1 &=||\alpha(\vx-\vy)+(\alpha-\beta)\vy ||_1 \\ 
				      & \leq \alpha||\vx - \vy||_1 +|\beta-\alpha| \,\, ||\vy||_1 \tag{triangle inequality}\\ 
				      & = \alpha ||\vx - \vy||_1 + \left|\frac{q}{X} - \frac{q}{Y}\right| Y \tag{definitions of $\alpha$, $\beta$, and $Y$} \\
				      & = \alpha ||\vx - \vy||_1 + \left|\frac{q}{X}Y - q\right|\\
				      & = \alpha ||\vx - \vy||_1 + \left|\frac{q}{X}Y - \frac{q}{X}X\right| \\
				      & = \alpha ||\vx - \vy||_1 + (1-\alpha)\left|Y - X\right| \tag{$1-\alpha = q/X$}\\
				      & \leq ||\vx - \vy||_1\tag{$|X-Y|\leq||\vx - \vy||_1$ by triangle inequality}
\end{align*}

\noindent \underline{Case 2:} $X < q$.\medskip

In this case we note that $Y \geq q > X >0$ ($X>0$ because $M<q$ and the seating of $w$ was plausible according to $P$) and hence $|Y-X|= Y-X$.
We split $X$ up into its ``negative part'' $X^-$ and ``positive part'' $X^+$ as follows:
\[
	X^- = \sum \{\max(0, -\vx(r) \colon r\in A_w)\} \; \text{ and } \; X^+ = \sum \{\max(0, \vx(r) \colon r\in A_w)\} .
\]
These definitions are such that $X = X^+- X^-$ and $||\vx||_1 = X^+ + X^-$. Now we have 
\begin{align}
	||\alpha \vx - \beta \vy ||_1 & \leq |\alpha| ||\vx||_1 + \beta ||\vy||_1 \notag \tag{triangle inequality} \\ 
				      & = |\alpha|(X^+ +X^-)+\beta Y.\label{eq:halfway}
\end{align}
At this point we note on the one hand that $\beta Y = Y-q$ is just the surplus of $w$ according to $P'$, and on the other hand 
\begin{align*}
	|\alpha|(X^+ + X^-) & = |\alpha|(X^+ - X^-) + 2|\alpha|X^- \\
			    & = q- X +2D_w.
\end{align*}
Putting this together, equation (\ref{eq:halfway}) reads as 
\begin{align*}
	||\alpha \vx - \beta \vy ||_1 & \leq |\alpha| ||\vx||_1 + \beta ||\vy||_1  \\ 
				      & = (q-X + 2D_w)+Y-q \\ 
				      & = Y-X +2D_w \\
				      & = |Y-X|+2D_w \\ 
					& \leq ||\vx-\vy||_1+2D_w.
\end{align*}

%
\bibliographystyle{splncs04}
\bibliography{bibliography.bib}

\end{document}